\documentclass[conference,a4paper]{IEEEtran}
\IEEEoverridecommandlockouts

\usepackage{cite}
\usepackage{amsmath,amssymb,amsfonts}
\usepackage{graphicx}
\usepackage{booktabs}
\usepackage{url}
\usepackage{xcolor}
\usepackage[hidelinks]{hyperref}

\begin{document}

\title{Do Rollups Self-Organize? Multi-Agent\\
Reinforcement Learning in the EIP-4844 Blob Fee Market}

\author{\IEEEauthorblockN{Alessio Brini}
\IEEEauthorblockA{\textit{Pratt School of Engineering} \\
\textit{Duke University} \\
Durham, NC, USA \\
alessio.brini@duke.edu}}

\maketitle

\begin{abstract}
EIP-4844 gave Ethereum rollups a separate fee dimension for data, governed by a multiplicative update rule that targets three ``blobs'' per block. Whether this target is reached depends on how many independent, self-interested rollups choose to post data, a question that closed-form models address only for a single agent under exogenous prices or in a static equilibrium. We propose a calibrated multi-agent reinforcement learning environment for the blob fee market in which the blob base fee is endogenous, set by the EIP-4844 rule on the aggregated decisions of heterogeneous rollups fit to post-Dencun Ethereum data. After confirming that the environment reproduces the known single-agent analytical optimum, we show that independent learners, with no convergence guarantee, still settle near the target without coordination ($3.32\pm0.01$ blobs per block over five seeds). Over a fiftyfold demand range ($10\times$ to $500\times$), supply stays at the target from $10\times$ to $200\times$ and rises above it at higher demand, where the fee can no longer hold it. Posting frequency increases with rollup size, an emergent analog of the size threshold predicted by prior theory. A costless layer-1 calldata outside option pushes the fee down by roughly an order of magnitude while aggregate supply stays at the target, so the fee rule acts as a negative-feedback controller on decentralized supply.
\end{abstract}


\section{Introduction}
\label{sec:intro}

The Dencun upgrade (March 2024) activated EIP-4844 (Ethereum Improvement Proposal 4844), which gave Ethereum rollups a dedicated data channel, \emph{blobs}, priced on a fee dimension separate from ordinary gas. The blob base fee follows a multiplicative update rule designed to hold realized usage at a target of three blobs per block. From a resource-market standpoint this is a control system in which the price steers aggregate usage back to the set point. Whether the set point is reached, however, is not a property of the rule alone. It depends on how many independent rollups, each minimizing its own posting and delay cost, respond to the price they collectively create.

Existing theory characterizes a single rollup's optimal posting policy under an \emph{exogenous} price process~\cite{baron2023, wang2025, mamageishvili2022}, and field studies document blob-market behavior empirically~\cite{park2024, heimbach2025, huang2024, lee2024}. Neither line answers the equilibrium question directly. Once the fee is endogenous in the agents' joint actions, the single-agent optimality arguments no longer apply. Relatedly,~\cite{crapis2023} characterize the choice between blobs and calldata, together with blob cost-sharing, in a static market equilibrium, but closed-form equilibrium strategies for the dynamic multi-agent regime with an endogenous fee are lacking. We take a computational route and ask whether the protocol's intended operating point \emph{emerges} when self-interested rollups learn how to post.

We make three contributions.
\begin{itemize}
  \item We propose a calibrated multi-agent reinforcement learning (RL) environment for the EIP-4844 blob market in which the blob base fee is endogenous, set by the update rule on the aggregated blob decisions of $N$ heterogeneous rollups whose arrival rates and price processes are fit to post-Dencun Ethereum data (Sections~\ref{sec:model} and~\ref{sec:calibration}).
  \item We show that independent proximal policy optimization (PPO)~\cite{schulman2017} learners self-organize to a steady state near the protocol target without any coordination mechanism. Posting frequency rises monotonically with rollup size, an endogenous analog of the arrival-rate threshold of~\cite{crapis2023}, and the empirical steady state stays at the target across a twentyfold range of aggregate demand (phase diagram, Section~\ref{sec:results}).
  \item We show that adding a costless layer-1 (L1) calldata\footnote{Calldata is the pre-EIP-4844 channel for rollup data, posted in the payload of an L1 transaction. It remains available and is cheaper than blobs only when the blob fee is well above its floor.} outside option collapses the fee level by roughly an order of magnitude while leaving aggregate supply pinned at the target, so the rule reaches its set point independently of the price level.
\end{itemize}

The environment also reproduces, as a sanity check, the single-agent analytical threshold of~\cite{baron2023}. All code and run configurations are at \url{https://github.com/Alessiobrini/eip4844-blob-marl}, including the calibrated parameters and the BigQuery queries that regenerate the raw on-chain data.

\section{Related Work}
\label{sec:related}

\textbf{Posting strategies under exogenous prices.} In the single-agent setting,~\cite{baron2023} derive threshold publishing policies under a broad class of price dynamics that includes martingales, and~\cite{wang2025} extend this analysis to fee-elastic Poisson arrivals under a mean-reverting L1 gas price, where the optimal queue threshold depends on the gas price. Earlier,~\cite{mamageishvili2022} design rollup batch-posting algorithms with operations-research tools. All assume the rollup is a price taker. Our environment makes the blob fee endogenous in the agents' joint actions.

\textbf{Blob-market economics and empirics.} \cite{crapis2023} analyze EIP-4844 economics and predict an arrival-rate threshold below which a (smaller) rollup should defect to calldata. Field studies~\cite{park2024, heimbach2025} characterize the early blob fee market, while~\cite{huang2024} document builder suboptimality and~\cite{lee2024} documents the unrealized potential of blob sharing. We complement these by treating the market as a learning system and asking whether its steady state is the one the protocol intends.

\textbf{Learning in fee markets.} Multi-agent RL with independent learners~\cite{tan1993} and agent-based market simulators~\cite{byrd2020abides} are established tools for studying emergent behavior in markets. The blob fee adopts the EIP-1559 congestion-pricing design (a base fee that adjusts toward a target, here in fully exponential form rather than the linear approximation used for gas), whose economics~\cite{roughgarden2021} and convergence dynamics~\cite{leonardos2021} are well studied. We apply those tools to a resource market with a protocol-defined target, asking whether the set point is reached.

\section{Model and Environment}
\label{sec:model}

One time step is one L1 block ($\approx 12$\,s). The market hosts $N$ heterogeneous rollups. Rollup $i$ maintains a queue $Q_{i,t}$ that grows each block by a Poisson draw with mean $\lambda_i$ and observes the local state
\begin{equation}
  s_{i,t} = \bigl(Q_{i,t}/\tau_i,\; B_t,\; P_t^{\mathrm{gas}}\bigr),
\end{equation}
where $B_t$ is the blob base fee (wei), $P_t^{\mathrm{gas}}$ the L1 gas base fee, and $\tau_i$ the rollup's transactions-per-blob capacity ($\tau_i = 2^{17}/d_i \approx 262$ for the homogeneous $d_i=500$ bytes per transaction we use). Dividing the queue by $\tau_i$ expresses it in blob units, keeping the observation order-one under arrival-rate scaling. The two price components enter in log form. We otherwise leave the observation unstandardized, with advantages normalized per update. Each rollup selects an action $a_{i,t}\in\{0,1,\dots,6\}$, the number of blobs to post in block $t$ ($0=\text{wait}$). A post of $a_{i,t}$ blobs removes $\min(a_{i,t}\,\tau_i,\,Q_{i,t})$ transactions from the queue. An underfilled blob still occupies, and pays for, a full blob.

\textbf{Endogenous blob base fee.} After every block the fee updates through the EIP-4844 rule\footnote{The spec accumulates excess blob gas across blocks and sets the fee to $\mathrm{MIN}\cdot\exp(\text{excess}/F)$. Equation~\eqref{eq:fee} is the per-block reduction of that recursion, and its denominator $8b^*=24$ differs by ${\sim}6\%$ from the spec's effective $F/2^{17}\approx25.5$, without affecting the qualitative results.} applied to the \emph{aggregate} blobs posted by all rollups,
\begin{equation}
  B_{t+1} = B_t \cdot \exp\!\Bigl(\frac{b_t - b^*}{8\,b^*}\Bigr),
  \qquad b_t = \sum_{i=1}^{N} b_{i,t},
  \label{eq:fee}
\end{equation}
with $b_{i,t}$ the blobs rollup $i$ posts ($a_{i,t}$ capped by queue availability), target $b^*=3$, and cap $b^{\max}=6$.\footnote{There is no builder-side inclusion rule. When aggregate requests exceed $b^{\max}$, the fee update counts $\min(b_t,b^{\max})$ while queues drain and costs accrue in full. In rollouts of the trained policies, the cap binds on one block in $10^5$ at $100\times$ demand, on $0.03\%$ of blocks at $200\times$, and on $8.4\%$ (mean excess $1.7$ blobs) at $500\times$, so this simplification affects only the saturation numbers.} Equation~\eqref{eq:fee} couples the rollups, so no agent sees the fee as exogenous. The L1 gas base fee follows a mean-reverting first-order autoregressive (AR(1)) process fit to data (Section~\ref{sec:calibration}).

\textbf{Cost and reward.} A rollup pays a delay cost on its standing queue and a posting cost for the blob gas and execution gas it consumes,
\begin{equation}
  r_{i,t} = -\bigl(\underbrace{\alpha_i\,Q_{i,t}}_{\text{delay}}
  + \underbrace{B_t\,b_{i,t}\cdot 2^{17}
  + P_t^{\mathrm{gas}}\,g^{\mathrm{fix}}}_{\text{posting}}\bigr),
  \label{eq:reward}
\end{equation}
where $2^{17}=131{,}072$ is the blob gas per blob and $g^{\mathrm{fix}}$ the fixed execution gas of a posting transaction. Delay (dimensionless queue) and posting (wei) costs live in different units. We reconcile them by a revealed-preference calibration of $\alpha_i$ that normalizes both by a reference posting cost $C^{\mathrm{ref}}_i = \bar{P}^{\mathrm{gas}}g^{\mathrm{fix}}$, giving a dimensionless effective delay weight $\tilde\alpha_i = \alpha_i/C^{\mathrm{ref}}_i = 2\lambda_i$.

\textbf{Outside option.} In the calldata variant (Section~\ref{sec:results}) a rollup may instead drain its full queue through a single L1 calldata transaction at cost $P_t^{\mathrm{gas}}(g^{\mathrm{fix}} + 16\,d_i\,Q_{i,t})$. Calldata uses no blob gas, so it does not enter~\eqref{eq:fee}. In our runs both execution-gas terms, here and in~\eqref{eq:reward}, are negligible, so calldata is effectively costless.

\textbf{Learners.} Each rollup runs an independent PPO policy, a multilayer perceptron actor-critic with two hidden layers of 64 units and $\tanh$ activations, learning rate $3\times10^{-4}$, clip coefficient $0.2$, generalized advantage estimation with $\lambda_{\mathrm{GAE}}=0.95$, discount $\gamma=0.99$, rollouts of $2{,}000$ blocks per update with four epochs of minibatches of $256$, value and entropy loss coefficients $0.5$ and $0.01$, and gradient-norm clip $0.5$. The implementation is PyTorch, and each $10^5$-step run trains in about a minute on a laptop CPU. There is no shared critic, communication, or reward sharing.

\section{Calibration}
\label{sec:calibration}

We calibrate on block-level Ethereum data from Google BigQuery's \texttt{crypto\_ethereum} dataset over the post-Dencun window (March 13 to August 31, 2024), covering $1{,}218{,}246$ blocks. We fit the L1 gas base fee and the log blob base fee each as an AR(1) process. The log parameterization for the blob fee matches the multiplicative structure of~\eqref{eq:fee}. Table~\ref{tab:calib} reports the fits. Both processes are highly persistent, the blob fee far more so.

We estimate per-rollup arrival rates $\lambda_i$ from type-3 transactions grouped by batcher address, counted over March 2024 to March 2025 and divided by the window's block count. These are L1 posting events, not the underlying layer-2 transaction arrivals, which for large rollups run to tens or hundreds per block. The calibrated $\lambda_i$ is thus a posting intensity. It enters the simulation as the mean of the Poisson arrivals feeding $Q_{i,t}$ and pins down relative sizes, while the demand level is set by the multiplier below, so results are read per demand regime. We label 18 rollups covering $86.3\%$ of post-Dencun blob-posting transactions. Their rates are well described by $\lambda_i\sim\mathrm{LogNormal}(-2.78,\,1.18)$ (Kolmogorov--Smirnov $p=0.98$). The five largest, namely Taiko, Base, Arbitrum One, Scroll, and World Chain (rates in Table~\ref{tab:postfreq}), anchor the $N=5$ experiments.\footnote{Taiko is a based rollup posting one blob per transaction by design. Taiko, World Chain, and Zircuit (18-rollup roster) launched mid-window, so their rates also average over blocks before launch. The target/cap are the Dencun-era values. Section~\ref{sec:discussion} scopes the results to this historical configuration.} We use these five for the per-rollup analysis and confirm generality on the 18-rollup roster below.

\begin{table}[t]
\centering
\caption{Calibrated AR(1) price processes (March to August 2024). Half-life is in blocks.}
\label{tab:calib}
\resizebox{\columnwidth}{!}{%
\begin{tabular}{lccccc}
\toprule
Process & $\mu$ & $\theta$ & $\sigma$ & $R^2$ & Half-life \\
\midrule
Gas fee (ETH/gas) & $1.09{\times}10^{-8}$ & $2.140{\times}10^{-3}$ & $9.85{\times}10^{-10}$ & $0.996$ & $324$ \\
Blob fee (log-wei) & $1.533$ & $1.242{\times}10^{-4}$ & $0.081$ & $0.9998$ & $5{,}582$ \\
\bottomrule
\end{tabular}}
\end{table}

\textbf{Demand scaling.} At empirical rates the top rollups in our roster post below the three-blob target (about $1.3$ blobs per block in simulation), so the fee sits at its floor and posting every block is trivially optimal, consistent with the sub-target demand and near-floor fees documented post-Dencun~\cite{park2024, heimbach2025}. The economically active regime is the forward-looking one, reached as rollup adoption grows, the growth that data-availability upgrades (Pectra, PeerDAS~\cite{eip7594}) are designed to accommodate. We therefore scale every $\lambda_i$ by a common multiplier, holding $\alpha_i$ at its empirical calibration, and treat the high-multiplier regime as a stress test of the fee rule, reporting the steady-state supply across a wide range of multipliers.

\section{Results}
\label{sec:results}

We train independent PPO learners to a steady state ($\sim$$10^5$ environment steps, $5\times10^4$ for the phase diagram and the target variation) over five random seeds and report each steady-state statistic as a mean $\pm$ s.d.\ across seeds, computed over the final $20\%$ of rollouts.\footnote{We judge convergence by the ordinary-least-squares slope of the aggregate blobs-per-block series over this window, which is negligible relative to its mean.}

\textbf{Single-agent validation.} Before the multi-agent results we confirm the environment reproduces known single-agent behavior (Table~\ref{tab:validation}). A deep Q-network (DQN)~\cite{mnih2015dqn} agent\footnote{The DQN is a Q-network with two hidden layers of 64 units and rectified-linear-unit activations, learning rate $10^{-4}$, replay buffer $10^5$, batch $64$, target network refreshed every $1{,}000$ steps, $\varepsilon$-greedy decayed over the first $30\%$ of training, $\gamma=0.995$ ($0.99$ for the last two regimes of Table~\ref{tab:validation}).} matches the threshold policy of~\cite{baron2023} under a log-normal random-walk price, posts every block at the fee floor, and times the market under extreme congestion.

\begin{table}[t]
\centering
\caption{Single-agent validation (exogenous-price special case). The environment matches the analytical threshold policy and times the market only when it should.}
\label{tab:validation}
\resizebox{\columnwidth}{!}{%
\begin{tabular}{lcc}
\toprule
Regime & DQN posting freq. & Reference \\
\midrule
Random-walk price~\cite{baron2023} & $0.46$ & $0.48$ (threshold policy) \\
Calibrated fee floor & $1.00$ & $1.00$ (always-post opt.) \\
Extreme congestion & $0.64$ & $8.8\times$ cheaper (always-post) \\
\bottomrule
\end{tabular}}
\end{table}

\begin{figure*}[t]
\centering
\includegraphics[width=\textwidth]{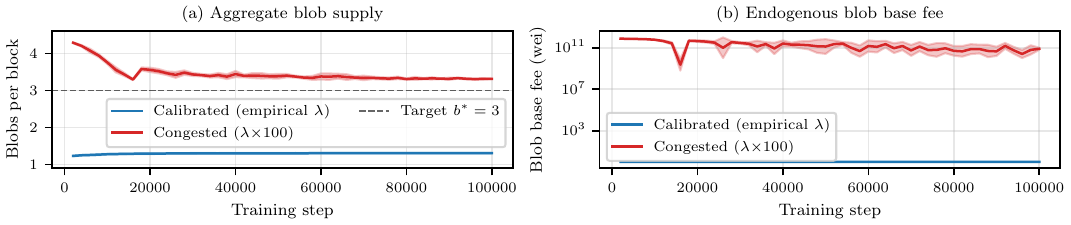}
\caption{Multi-agent training dynamics, $N=5$ (each curve is the mean across five seeds, shaded $\pm$ s.d.). (a) Aggregate blob supply against the EIP-4844 target $b^*=3$, at empirical arrival rates (blue) and under $100\times$ demand (red). (b) Endogenous blob base fee (log scale).}
\label{fig:conv}
\end{figure*}

\begin{figure}[!t]
\centering
\includegraphics[width=\columnwidth]{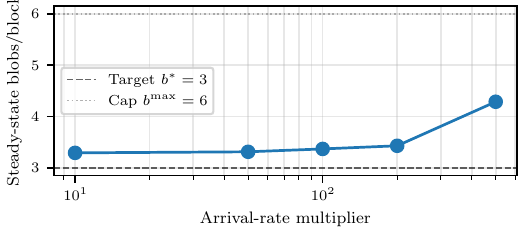}
\caption{Steady-state blob supply versus the arrival-rate multiplier, $N=5$ (mean $\pm$ s.d.\ over five seeds), against the target $b^*=3$ and the cap $b^{\max}=6$.}
\label{fig:phase}
\end{figure}

\begin{figure*}[!t]
\centering
\includegraphics[width=\textwidth]{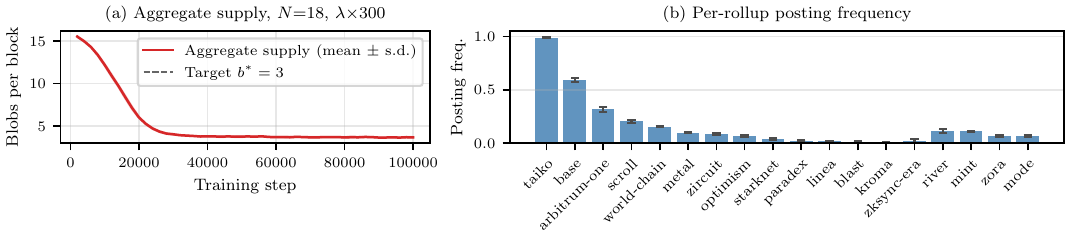}
\caption{Generality across roster size, with all 18 calibrated rollups at $300\times$ demand (mean over five seeds, shaded/error bars $\pm$ s.d.). (a) Aggregate supply against the target $b^*=3$. (b) Per-rollup posting frequency, ordered by decreasing $\lambda_i$.}
\label{fig:n18}
\end{figure*}

\begin{figure}[!t]
\centering
\includegraphics[width=\columnwidth]{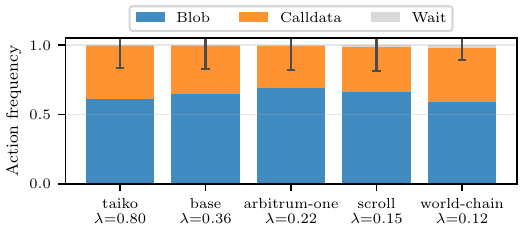}
\caption{Per-rollup action mix in the congested regime with a calldata outside option (mean over five seeds, error bars $\pm$ s.d.\ on the calldata fraction), ordered by decreasing $\lambda_i$.}
\label{fig:calldata}
\end{figure}

\textbf{Reaching the target.} Fig.~\ref{fig:conv} shows the calibrated and the congested regime for $N=5$. Under $100\times$ demand the dynamics become economically non-trivial. The five independent learners settle to $\mathbf{3.32\pm0.01}$ blobs per block, slightly above the protocol target $b^*=3$, and the fee holds on the order of $10^{10}$ to $10^{11}$\,wei, far above the $1$-wei floor but below the simulation's numerical fee ceiling. The endogenous feedback of~\eqref{eq:fee} alone drives the population to the set point, and the tight band across seeds rules out a lucky seed.

\textbf{Non-learning baselines.} Rule-based policies in the same congested environment (five seeds, $2\times10^4$ blocks) separate the fee rule's contribution from learning. Price-blind rules miss the target from both sides. Draining the queue every block yields $5.00$ blobs per block with the fee pinned at the numerical ceiling, and posting only full blobs yields $0.63$ with the fee at its floor. A myopic threshold that posts when the delay cost exceeds the posting cost at current prices, a rule that knows the cost model, also reaches the target ($3.29$, s.d.\ $<0.01$). The controller thus pins the aggregate whenever supply responds to price. Learning recovers the price response from rewards alone, at higher realized cost than the informed rule ($236$ to $715$ versus $221$ in normalized units).

\textbf{Strategy ordering in $\lambda_i$.} In the congested regime, posting frequency increases with the rollup's arrival rate (Table~\ref{tab:postfreq}). Averaged across seeds, the ordering is strictly monotone (and holds in four of five seeds individually). The largest rollup (Taiko) posts in $97\%$ of blocks, the smallest (World Chain) in $36\%$. Low-volume rollups thus wait out most blocks, shrinking their realized blob footprint. This is an endogenous, posting-frequency form of the calldata arrival-rate threshold of~\cite{crapis2023}, here without an explicit outside option. Since the delay weight $\tilde\alpha_i=2\lambda_i$ also scales with size, we retrain with a common $\tilde\alpha_i$ set from the roster-mean rate. Aggregate supply is nearly unchanged ($3.39\pm0.02$ versus $3.32\pm0.01$) and Taiko still posts most ($0.80\pm0.05$), but the gradient flattens ($0.80$ to $0.62$) and its ordering is not stable across seeds, so the steep gradient of Table~\ref{tab:postfreq} reflects arrival volume and the calibrated delay pricing jointly.

\begin{table}[t]
\centering
\caption{Learned posting frequency, $N=5$ congested regime ($100\times$ demand), mean $\pm$ s.d.\ over five seeds. Posting frequency is the fraction of blocks in which a rollup posts at least one blob. The aggregate (last row) is the mean total blobs per block measured directly from the runs (a posting rollup may place several blobs).}
\label{tab:postfreq}
\begin{tabular}{lcc}
\toprule
Rollup & $\lambda_i$ (tx/block) & Posting freq. \\
\midrule
Taiko        & $0.803$ & $0.97\pm0.02$ \\
Base         & $0.361$ & $0.84\pm0.06$ \\
Arbitrum One & $0.218$ & $0.67\pm0.07$ \\
Scroll       & $0.153$ & $0.47\pm0.05$ \\
World Chain  & $0.125$ & $0.36\pm0.03$ \\
\midrule
\multicolumn{2}{l}{Aggregate blobs/block} & $3.32\pm0.01$ \\
\bottomrule
\end{tabular}
\end{table}

\textbf{Phase diagram.} Fig.~\ref{fig:phase} varies the arrival-rate multiplier from $10\times$ to $500\times$ (fiftyfold) at $N=5$ (five seeds per point). The steady-state supply stays within one blob of the target for multipliers from $10$ to $200$ ($3.29$ to $3.43$ blobs per block), then rises to $4.29$ at $500\times$, where the fee approaches the simulation's numerical ceiling and can no longer price-ration supply to the target, the regime in which a strategic builder or an outside option becomes economically necessary. Scaling the roster to all 18 calibrated rollups (at $300\times$) yields the same picture (Fig.~\ref{fig:n18}). Aggregate supply settles at $3.66\pm0.01$ blobs per block, and posting frequency broadly increases with $\lambda_i$, strictly so among the 13 largest rollups, while the four smallest post more often than the next few.

\textbf{Calldata outside option.} Re-running the $N=5$ congested regime with the calldata action enabled (Fig.~\ref{fig:calldata}) separates an aggregate effect stable across seeds from a distributional one that is not. The stable effect is twofold. The fee level collapses by roughly an order of magnitude (geometric-mean fee $\sim$$10^{10}$\,wei with the option versus $\sim$$10^{11}$ without), while the \emph{aggregate} blob supply stays pinned near the target at $3.20\pm0.08$ per block (Equation~\eqref{eq:fee} drives the aggregate to target as long as some rollups use blobs). Which rollups defect varies with the seed. Across five seeds every rollup uses calldata $31$ to $39\%$ of the time, with no monotone size ordering, so we draw the size-threshold conclusion only from the posting-frequency gradient, not from calldata defection.

\section{Discussion and Limitations}
\label{sec:discussion}

\textbf{Feedback on the aggregate.} The target is an attractor of the fee dynamics. The exponent in~\eqref{eq:fee} depends only on the total $b_t=\sum_i b_{i,t}$, so the fee rises when aggregate posting exceeds $b^*$ and falls when posting is below it, whatever the posters' identity. The rule has a stable fixed point at $b^*$. That uncoordinated learners actually settle there, however the blobs are split, is the empirical content. The steady state sits slightly above $b^*$ ($3.32$ versus $3$) because the fee restarts at $1$\,wei every $2{,}000$-block episode, and by~\eqref{eq:fee} climbing back takes excess supply. The arrival-rate threshold surfaces here on the \emph{intensive} margin (Table~\ref{tab:postfreq}) rather than on the \emph{extensive} margin of a binary calldata defection (Fig.~\ref{fig:calldata}).

\textbf{Varying the target.} Holding demand fixed at $200\times$ and varying $b^*\in\{2,3,4\}$, the steady-state supply moves one-for-one with the target ($2.38\pm0.01$, $3.43\pm0.03$, and $4.47\pm0.02$), each $0.38$ to $0.47$ above target. So $b^*$ sets the operating point, while $8b^*$ in~\eqref{eq:fee} governs only adjustment speed.

\textbf{Protocol scope.} All experiments use the Dencun-era configuration ($b^*=3$, $b^{\max}=6$) in force over the calibration window. Pectra (May 2025) raised the pair to $6/9$~\cite{eip7691}, blob-parameter-only forks~\cite{eip7892} after the Fusaka upgrade reached $14/21$ in January 2026, EIP-7918 now bounds the blob base fee from below by execution cost~\cite{eip7918}, and EIP-7623 raised calldata costs~\cite{eip7623}. The one-for-one response of supply to the target above suggests the picture transfers to larger targets, but the calldata fee drop is specific to a market without a reserve price, and our calldata option is effectively costless.

Several limitations bound these claims. Convergence is empirical, not by theorem (no general guarantee for independent-learner multi-agent RL~\cite{hernandezleal2019}). The tight bands across seeds ($\pm0.01$) are our evidence. The L1 gas process is exogenous, and the builder is a rule-based pass-through that abstracts away builder-side strategy, empirically often suboptimal~\cite{huang2024}. The calldata option is effectively costless. Natural next steps are a learning builder and a cooperative blob-sharing mechanism.

\section{Conclusion}
\label{sec:conclusion}

We presented a calibrated multi-agent RL environment for the EIP-4844 blob fee market with an endogenous fee. Across a wide demand range, independent PPO learners self-organize to the protocol target without coordination, with posting frequency ordered by rollup size. A costless calldata outside option drops the fee level while supply holds at target. The fee rule thus acts as a self-stabilizing congestion signal for learning rollups, a property invisible to single-agent, exogenous-price analysis, and the calibrated environment is a reusable testbed for vetting fee-rule changes before deployment, with $b^*$ a clean design knob.

\bibliographystyle{IEEEtran}
\bibliography{references}

\end{document}